\documentclass[conference]{IEEEtran}
\IEEEoverridecommandlockouts

\usepackage{cite}
\usepackage{amsmath,amssymb,amsfonts}
\usepackage{algorithmic}
\usepackage{graphicx}
\usepackage{textcomp}
\usepackage{xcolor}
\usepackage{array}
\usepackage{eso-pic}
\newcolumntype{C}[1]{>{\centering\arraybackslash}p{#1}}
\def\BibTeX{{\rm B\kern-.05em{\sc i\kern-.025em b}\kern-.08em
    T\kern-.1667em\lower.7ex\hbox{E}\kern-.125emX}}
\AddToShipoutPictureBG*{
  \AtPageUpperLeft{
    \raisebox{-3\baselineskip}{
      \makebox[\paperwidth][c]{
        \parbox{0.9\textwidth}{\centering\footnotesize 
          © 2025 IEEE. Personal use of this material is permitted. Permission from IEEE must be obtained for all other uses, in any current or future media, including reprinting/republishing this material for advertising or promotional purposes, creating new collective works, for resale or redistribution to servers or lists, or reuse of any copyrighted component of this work in other works.
        }
      }
    }
  }
}
\begin{document}

\title{A High-Level Feature Model to Predict the Encoding Energy of a Hardware Video Encoder\\

}

\author{\IEEEauthorblockN{Diwakara Reddy\IEEEauthorrefmark{1}, Christian Herglotz\IEEEauthorrefmark{1}\IEEEauthorrefmark{2}, and André Kaup\IEEEauthorrefmark{1}}
\IEEEauthorblockA{\IEEEauthorrefmark{1}Chair of Multimedia Communications and Signal Processing,\\
Friedrich-Alexander-Universität Erlangen-Nürnberg (FAU) \\
\{diwakara.reddy.kadagathur, christian.herglotz, andre.kaup\}@fau.de}
\IEEEauthorblockA{\IEEEauthorrefmark{2}Chair of Computer Engineering, \\
Brandenburg University of Technology Cottbus-Senftenberg\\
}
}

\maketitle

\begin{abstract}
In today's society, live video streaming and user generated content streamed from battery powered devices are ubiquitous. Live streaming requires real-time video encoding, and hardware video encoders are well suited for such an encoding task. In this paper, we introduce a high-level feature model using Gaussian process regression that can predict the encoding energy of a hardware video encoder. In an evaluation setup restricted to only P-frames and a single keyframe, the model can predict the encoding energy with a mean absolute percentage error of approximately 9\%. Further, we demonstrate with an ablation study that spatial resolution is a key high-level feature for encoding energy prediction of a hardware encoder. A practical application of our model is that it can be used to perform a prior estimation of the energy required to encode a video at various spatial resolutions, with different coding standards and codec presets.  
\end{abstract}


\section{Introduction}
In the current decade, live streamed content and User Generated Content (UGC) are popular video content types \cite{Statista2024, Statista2025}. Live streaming necessitates real-time video encoding and usually relies on hardware video encoders. When UGC is created on handheld battery operated devices, it is important to perform energy conscious video encoding. Additionally, energy aware video encoding is important to reduce the carbon footprint of video streaming \cite{shiftproject}.

There exist two implementation types of video encoders, namely software (SW) and hardware (HW). SW encoders are designed to run on general
purpose processors, which allows portability across different machines. Performance of a SW encoder is dependent on the computational resources of the host machine. Slow presets of a SW encoder offer high compression efficiency at the cost of encoding speed, the faster presets trade-off compression gains for speed and offer low latency encoding. HW encoders run on dedicated Application Specific Integrated Circuits (ASICs), which provides capability for accelerated and energy efficient encoding, they offer real-time encoding at the cost of compression efficiency.

A literature review suggests that there are many models for predicting the energy demand of SW video encoders. In \cite{Katsenou2022,Chachou2024}, the authors discuss the energy efficiency of various state-of-the-art video codecs. However, they do not provide a model to estimate the energy consumption. In \cite{Rodríguez2015}, the authors present an encoding energy and time model for an H.265 encoder. However, the model is only valid for the All Intra (AI) coding configuration. In \cite{Ramasubbu2022ICIP,Ramasubbu2022PCS,Ramasubbu2024}, the authors address the drawbacks of \cite{Rodríguez2015} and present more comprehensive and accurate models. Nevertheless, the models only predict the energy of a H.265 SW video encoder. Eichermüller et al. in \cite{Eichermüller2024} provide an encoding time and energy model for the SVT-AV1 video codec. SVT-AV1 is a SW implementation of the AV1 standard from Alliance for Open Media (AOM). Lachini et al. in \cite{Lachini2024} provide a framework for energy and $\text{CO}_{2}$ emissions estimation in the context of a cloud based video encoding. Their model is robust to include SW implementations of H.264 and H.265 encoders, however they only provide results for the medium presets.

There is limited research available on the energy consumption prediction of HW video encoders \cite{Amaral2016,Gan2007}. Still, there is a substantial research focussed on the energy prediction of HW video decoders \cite{Herglotz2018ISCAS, Herglotz2018,Kränzler2023,KränzlerPhDThesis}. Herglotz et al. in \cite{Herglotz2018ISCAS} introduced a High-Level (HL) feature model to estimate the energy of a H.265 HW decoder. Extending the work of \cite{Herglotz2018ISCAS}, Kränzler proposes separate models to estimate the energy of HW decoder implementations of H.264, H.265, VP9, and AV1 coding standards in \cite{KränzlerPhDThesis}. In this work, we introduce a HL feature model using Gaussian Process Regression (GPR) that can predict the encoding energy of a HW video encoder with a Mean Absolute Percentage Error (MAPE) of 9.08\%. 

Our contributions in this paper extends and addresses the gaps in existing knowledge as follows: (1) We extend the HL feature model in \cite{Herglotz2018ISCAS} for a HW decoder to a HW encoder, (2) The HL model for decoder energy prediction cannot be directly ported to perform prior estimation of encoder energy, because bitstream size is one of the HL features in \cite{Herglotz2018ISCAS}. This information is readily available for a decoder, but not for an encoder. We address this by modifying the HL feature model to include only the features that are available before encoding, (3) In lieu of a separate model per standard for HW decoder energy presented in \cite{KränzlerPhDThesis}, we propose a single model for HW encoder energy prediction that considers three different standards and two encoder presets. We have organized this paper as follows: Section \ref{sec:Setup} presents details on the HW encoder used in our experiments, energy measurements, and energy modelling. Section \ref{sec:Results} discusses the modelling results and examines the relationship between various HL features and encoding energy. Finally, Section \ref{sec:Conclusion} concludes the paper.

\section{Measurement Setup and Modelling}
\label{sec:Setup}
\begin{figure}
    \includegraphics{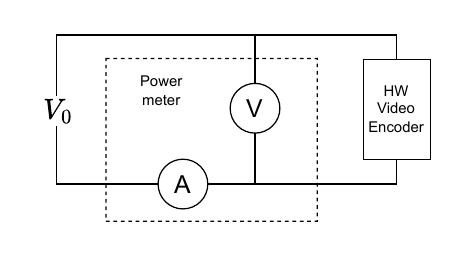}
    \caption{Energy measurement setup. $V_{0}$ is an AC voltage source.}
     \label{fig:EnergyMeasurementSetup}
\end{figure}

We use the NVIDIA Jetson Orin NX development kit\cite{JetsonOrinNX} as the HW encoder. It is powered by an ARM Cortex-A78E processor which is built on aarch64 architecture and features 16GB of RAM. It provides hardware-accelerated encoding support for H.264, H.265, and AV1 video coding standards. It offers four presets, namely, \textit{ultrafast}, \textit{fast}, \textit{medium}, and \textit{slow}. Encoding is done with the \textit{video\_encode} module, which is part of the NVIDIA Jetson Multimedia API. The device is connected to the Internet via ethernet and encoding is performed through remote access from a workstation. The development kit is connected to the ZES Zimmer LMG611 powermeter as shown in Fig. \ref{fig:EnergyMeasurementSetup} to perform energy measurements. Following the methodology to measure decoding energy in \cite{Herglotz2018}, we measure encoding energy $E_{\mathrm{enc}}$ as a difference between two consecutive energy measurements $E_{\mathrm{dynamic}}$ and $E_{\mathrm{static}}$ 
\begin{equation}
    E_{\mathrm{enc}} = E_{\mathrm{dynamic}} - E_{\mathrm{static}}
    \label{eq:EncodingEnergyEquation}
\end{equation}
$E_{\mathrm{dynamic}}$  and $E_{\mathrm{static}}$ are defined as
\begin{equation}
    E_{\mathrm{dynamic}} = \int_{t_0}^{t_0 + T} P_{\mathrm{dynamic}}(t) dt
\end{equation}
\begin{equation}
    E_{\mathrm{static}} = \int_{t_1}^{t_1 + T} P_{\mathrm{static}}(t) dt \text{,}
\end{equation}
where $P_{\mathrm{dynamic}}$ is the power consumption during the encoding process, $P_{\mathrm{static}}$ is the power consumption during the idle mode, $T$ is the encoding time, and $t_0$ and $t_1$ are two subsequent time instants. Measurement of energy can be a noisy process. To increase the statistical validity of the measured energy values, we perform Confidence Interval Tests (CITs) as explained in \cite{Herglotz2018, Bendat1971}. The test condition is defined as
\begin{equation}
    \Delta c  < \beta \cdot \overline{E_{\mathrm{enc}}}
\label{eq:CICondition}
\end{equation}
and 
\begin{equation}
    \Delta c =  2 \cdot \frac{\sigma}{\sqrt{m}} \cdot t_{\alpha}(m-1) \text{,}
\end{equation}
where $\beta$ represents the acceptable deviation of the measured encoding energy from the true encoding energy, $\overline{E_{\mathrm{enc}}}$ is the arithmetic mean of energy measurements, $m$ denotes the number of measurements, $\sigma$ indicates the standard deviation of the measured values, and $t_{\alpha}$ represents the student's t-distribution. We set $\alpha$  to 0.99 and $\beta$ to 0.02 based on \cite{Herglotz2018}. We stop energy measurements for a particular video sequence when the condition in (\ref{eq:CICondition}) is met. We then use the arithmetic mean of the measured energy values as the encoding energy $E_{\mathrm{true}}$ for the particular video sequence.

 Table \ref{tab:HLFeatureList} lists the HL features used for modelling the encoding energy. Modelling of the energy is done with GPR \cite{Rasmussen2006} based on the work of \cite{KränzlerPhDThesis, Kränzler2024arxiv}. GPR is a probabilistic supervised machine learning algorithm. It has the capability to account for measurement noise, hence it is well suited for our scenario. In \cite{KränzlerPhDThesis, Kränzler2024arxiv}, the author models decoding energy and demonstrates that GPR can provide a better prediction performance in comparison to Linear Regression (LR). In the presence of measurement noise, a linear regression model to predict encoding energy can be written as \cite{Rasmussen2006}
\begin{equation}
    \hat{E}_{enc} = \mathbf{x}^T\mathbf{w} + \epsilon \text{,}
\end{equation}
where $\mathbf{x}$ represents features $\mathrm{x}_0$-$\mathrm{x}_8$ in Table \ref{tab:HLFeatureList} , $\mathbf{w}$ indicates the weights, and $\epsilon$ is the noise. For our modelling, we assume $\epsilon$ is an independent identically distributed Gaussian noise of mean $0$ and variance $\sigma^2_n$, which is represented as \cite{Rasmussen2006}

\begin{equation}
    \epsilon \sim \mathcal{N} (0, \sigma^2_n) \text{.}
\end{equation}

A function approximator modelled by GPR can be represented as \cite{Rasmussen2006}
\begin{equation}
    f(\mathbf{x}) \sim \mathcal{GP}(m(\mathbf{x}), \Sigma) \text{,}
\end{equation}
where $m(\mathbf{x})$ indicates the mean function, and $\Sigma$ represents the covariance function. If we model the mean function with a basis function $b(\mathbf{x})$, then $f(\mathbf{x})$ can be modelled with a zero mean Gaussian process given by
\begin{equation}
    f(\mathbf{x}) \sim  b(\mathbf{x}) + \mathcal{GP}(0 , \Sigma) \text{.}
\end{equation}
Further, we approximate the covariance function with a kernel function. In our case, we perform modelling with the fitrgp function in Matlab \cite{fitrgpmatlab} with a linear basis function and an exponential kernel function. If $x_p$ and $x_q$ are two input features, the kernel function to calculate the co-variance between them is defined as \cite{KränzlerPhDThesis}
\begin{equation}
k(x_p,x_q) = \sigma^2_f \exp \left(-\frac{| x_p - x_q|}{l}\right) + \sigma^2_n \cdot \delta_{st}   \text{,} 
\end{equation}
where $\sigma^2_f$ denotes variance of the function $f(\mathbf{x})$, $l$ indicates the characteristic length scale, $\sigma^2_n$ denotes variance of the noise, and $\delta_{st}$ represents the Kronecker delta. To summarize, if $h(\mathbf{x})$ represents a set of linear basis functions, 
the model output is given by
\begin{equation}
    \hat{E}_{enc} = h(\mathbf{x})^T\beta + g(\mathbf{x}) \text{,}
\end{equation}
where $g(\mathbf{x}) \sim  \mathcal{GP}(0, \Sigma )$. The parameters $\beta$, $\sigma^2_f$, $l$, and $\sigma^2_n$ are inferred from data in the training phase. To account for overfitting, we perform 10-fold cross validation during the training process.
\begin{table}
\centering
\caption{Model Features. Features $\mathrm{x}_3$-$\mathrm{x}_7$ are boolean features that are set to 1 based on the standard and preset chosen for encoding. Feature $\mathrm{x}_0$ is a bias term, which is always set to 1}
\begin{tabular}{C{1cm}|C{6cm}} % Defines column widths
\hline
\textbf{Identifier} & \textbf{Feature} \\
\hline
$\mathrm{x}_0$ & offset energy \\
$\mathrm{x}_1$ & number of encoded frames \\
$\mathrm{x}_2$ & number of pixels (width $\times$ height) \\
$\mathrm{x}_3$ & standard\_H264 \\
$\mathrm{x}_4$ & standard\_H265 \\
$\mathrm{x}_5$ & standard\_AV1 \\
$\mathrm{x}_6$ & preset\_ultrafast \\
$\mathrm{x}_7$ & preset\_slow \\
$\mathrm{x}_8$ & QP \\
\end{tabular}
\label{tab:HLFeatureList}
\end{table}

\begin{table}
\centering
\caption{QPs chosen for different coding standards}
\begin{tabular}{C{2.5cm}|C{2.5cm}} % Defines column widths
\hline
\textbf{H264 and H265} & \textbf{AV1} \\
\hline
22, 27, 32, 37 & 108, 132, 160, 184
\end{tabular}
\label{tab:QPRange}
\end{table}

\begin{figure}[t]
    \includegraphics[width=0.5\textwidth]{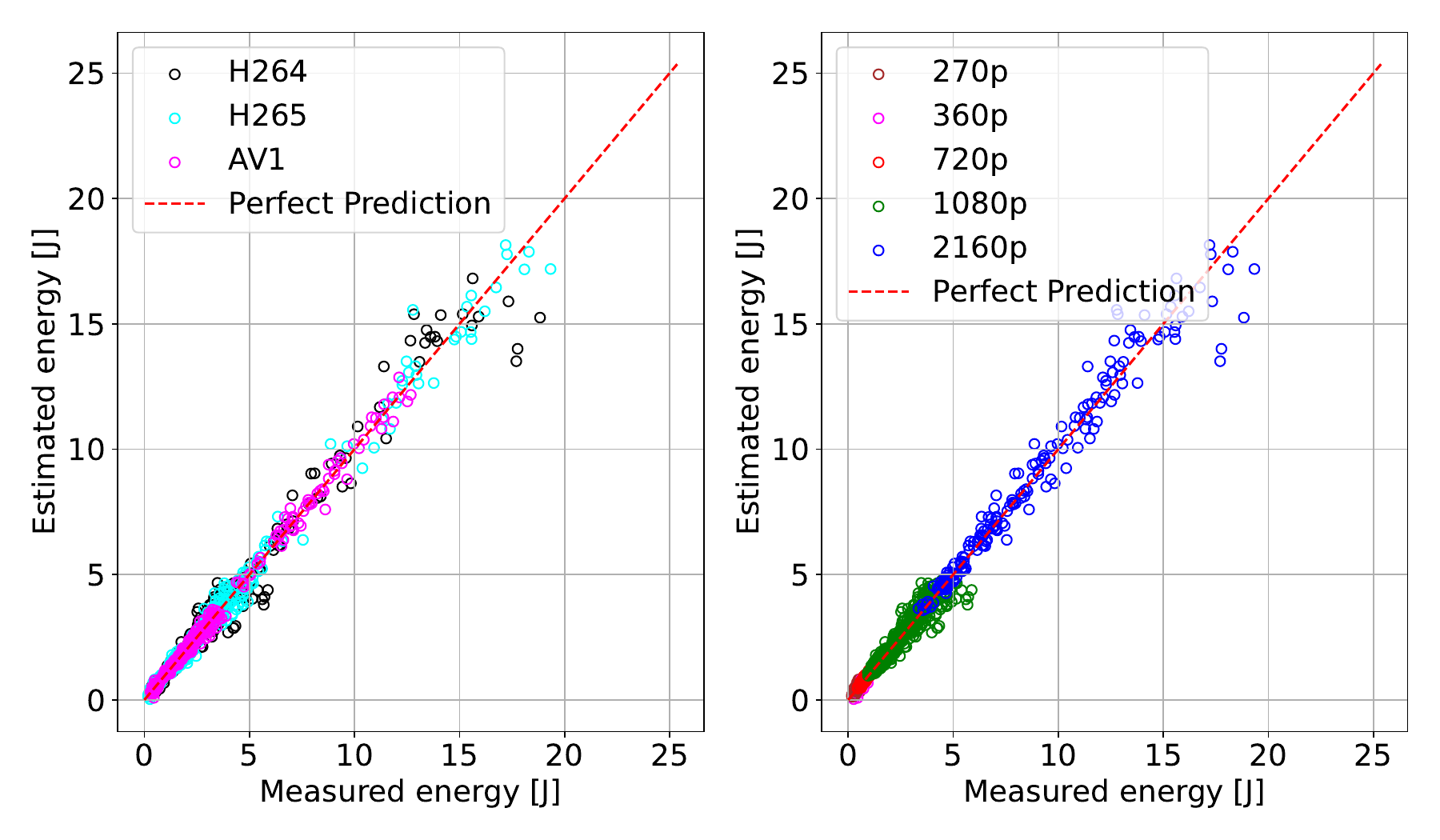}
    \caption{Visualization of modelling results. (left) Grouped by coding standard. (right) Grouped by vertical spatial resolution.}
    \label{fig:ModellingResults}
\end{figure}

\begin{figure}
    \includegraphics[width=0.5\textwidth]{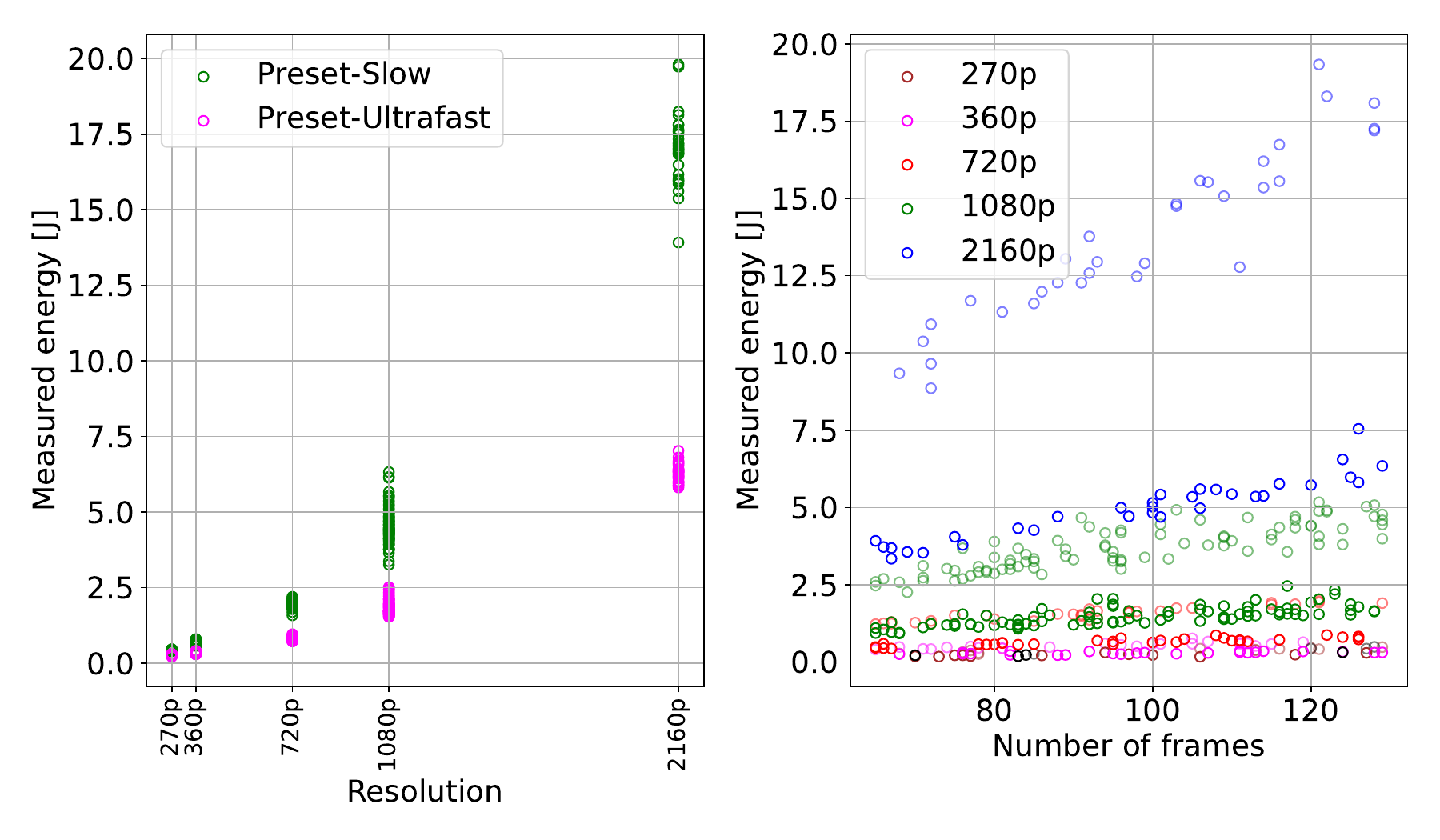}
    \caption{(left) Encoding energy consumption versus vertical spatial resolution. (right) Encoding energy versus number of frames, lighter markers correspond to the \textit{slow} preset, while darker markers indicate the \textit{ultrafast} preset. }
    \label{fig:EnergyvsFeatures}
\end{figure}

\section{Evaluation}
\label{sec:Results}
We present the modelling results for natural video sequences from classes A1-A5 of AOM Common Test Conditions (CTC) \cite{AOMCTCV2}. The test set includes 270p, 360p, 720p, 1080p, and 2160p (4K) video sequences. The HW encoder used in our experiments supports encoding of only 8-bit sequences, hence we convert 10-bit input sequences in the CTC to 8-bit sequences. The number of frames for encoding is chosen randomly between 65 and 130 for each sequence with a single keyframe similar to a low delay intra-frame refresh strategy. We perform modelling for H264, H265, and AV1 standards with no B-frames. The HW encoder provides the capability to explicitly specify the number of B-frames as an input argument, for our experiments however, we use the default configuration which has no B-frames. Additionally, we consider presets \textit{ultrafast} and \textit{slow} for energy modelling. Only these two presets are considered as our experiments indicate that presets \textit{fast}, \textit{medium}, and \textit{slow} have identical rate-distortion performance. We use constant QP as the rate-control method. The HW encoder allows a QP range of 0-51 for H264 and H265, and 1-255 for AV1. Considering the QP mapping between SVT-AV1 and the AV1 standard, and based on the previous work in \cite{Katsenou2022,JVET2020}, we use QPs
listed in Table \ref{tab:QPRange}. 

Accuracy is measured in terms of MAPE which is defined as
\begin{equation}
    MAPE = \frac{1}{B} \sum_{i=1}^{B} \frac{|E_{\mathrm{true},i} - E_{\mathrm{est},i}|}{E_{\mathrm{true},i}} \times 100 \text{,}
\label{eq:MAPE}
\end{equation}
where $B$ is number of bitstreams, $E_{\mathrm{true},i}$ and $E_{\mathrm{est},i}$ are measured and estimated energies. Considering that we perform 10-fold cross validation, each bitstream is part of the training set 9 times and the validation set once. $E_{\mathrm{est},i}$ is recorded when a bitstream is part of the validation set, this is then used to determine MAPE. Our model achieves an MAPE of 9.08\%. 

Fig. \ref{fig:ModellingResults} shows the visual representation of prediction and true energy. In this plot and later plots, each marker corresponds to one bitstream. We can notice that in most cases, the predicted value is close to the true value. In the right plot, we observe clusters corresponding to different resolutions, suggesting a dependency of the encoder energy on the video resolution (or number of pixels). When plots with resolution information are presented, the resolution corresponds to vertical resolution. However for the portrait sequences in the CTC, we group them according to their horizontal resolution.  

\begin{figure}
    \includegraphics[width=0.5\textwidth]{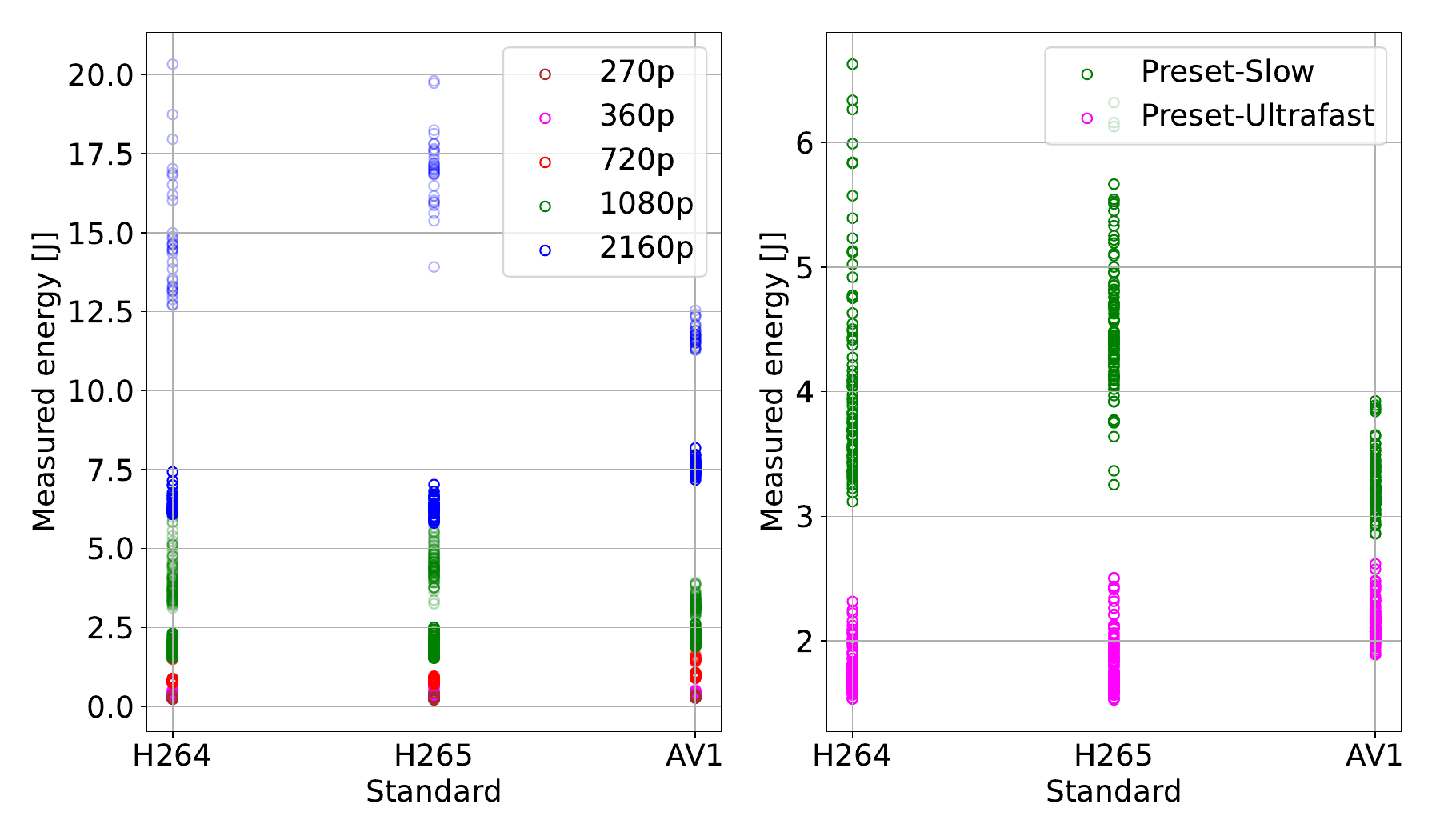}
    \caption{(left) Encoding energy consumption versus coding standard, lighter markers correspond to the \textit{slow} preset and darker ones to the \textit{ultrafast} preset. (right) Same plot with only 1080p resolution sequences.}
    \label{fig:EnergyvsCodec}
\end{figure}

\begin{figure}
    \includegraphics[width=0.5\textwidth]{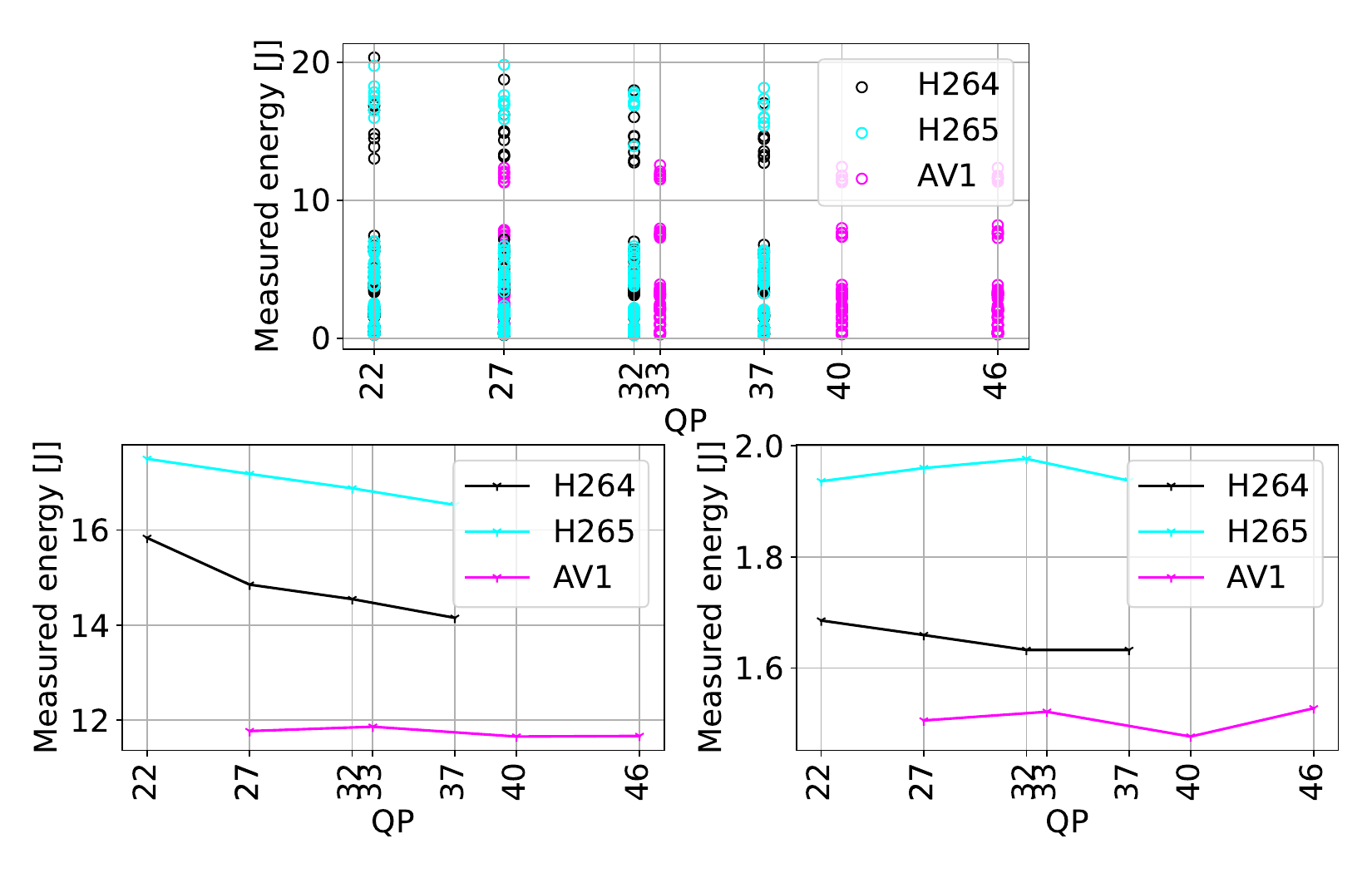}
    \caption{(top) Encoding energy versus QP values. (bottom left) Same plot with arithmetic mean of encoding energy across bitstreams for only 4K sequences and \textit{slow} preset, and (bottom right) only 720p sequences and \textit{slow} preset. QP values of the AV1 standard are divided by 4 in the plots.}
    \label{fig:EnergyvsQP}
\end{figure}

Fig. \ref{fig:EnergyvsFeatures} shows the relation between encoding energy and resolution and number of frames, only for the H.265 standard to facilitate clarity and interpretability. We can notice a correlation between encoding energy and resolution in the left plot and a correlation between encoding energy and number of frames in the right plot. The correlation shows that our approach to include spatial resolution and number of frames as features is a reasonable approach. The number of encoded frames is set to 130 for all the video sequences to generate the plots in Fig. \ref{fig:EnergyvsCodec} and Fig. \ref{fig:EnergyvsQP} and the left plot in Fig. \ref{fig:EnergyvsFeatures}. Fig. \ref{fig:EnergyvsCodec} shows the relation between energy consumption and coding standard. The difference in the energy consumption for 4K videos is easily noticeable in the left plot, however the differences for other resolutions is not evident. We present the data only for 1080p resolution in the right plot. It can be noticed in the right plot that there is only minor variation in the energy consumption for the \textit{ultrafast} preset across the three standards, however there is a more noticeable variation in the encoding energy for the \textit{slow} preset across the three standards. Fig. \ref{fig:EnergyvsQP} illustrates the relationship between energy consumption and the QP value. The top plot presents the data for all the bitstreams together, however restricting the data to a single resolution and preset in the bottom plots demonstrate that the correlation between QP value and encoding energy is dependent on the standard and resolution. We notice a monotonic relationship between QP value and energy for H.264 and H.265 in the bottom left plot, however that relationship is not maintained for 720p sequences in the bottom right plot. Furthermore, we observe no noticeable correlation between the encoding energy and the QP value for AV1. 

We performed training on a notebook running Windows 11 operating system and fitted with an Intel i5-10210 processor and an integrated Intel graphics processor, and 8GB of RAM. Training and validation of the model took 21.25 seconds and 3.7 milliseconds, respectively. We performed CITs with the same parameters as stated in the previous section to measure the training and validation times. Training time includes the time required for training and validation of all 10 folds in the 10-fold cross validation. However, the time required for validation of each fold in the 10-fold validation is presented as validation time. Validation time is an indicator for the inference time. The reasonable training and validation times indicate that the model is lightweight and does not require special HW such as a GPU.

\begin{figure}
    \includegraphics[width=0.5\textwidth]{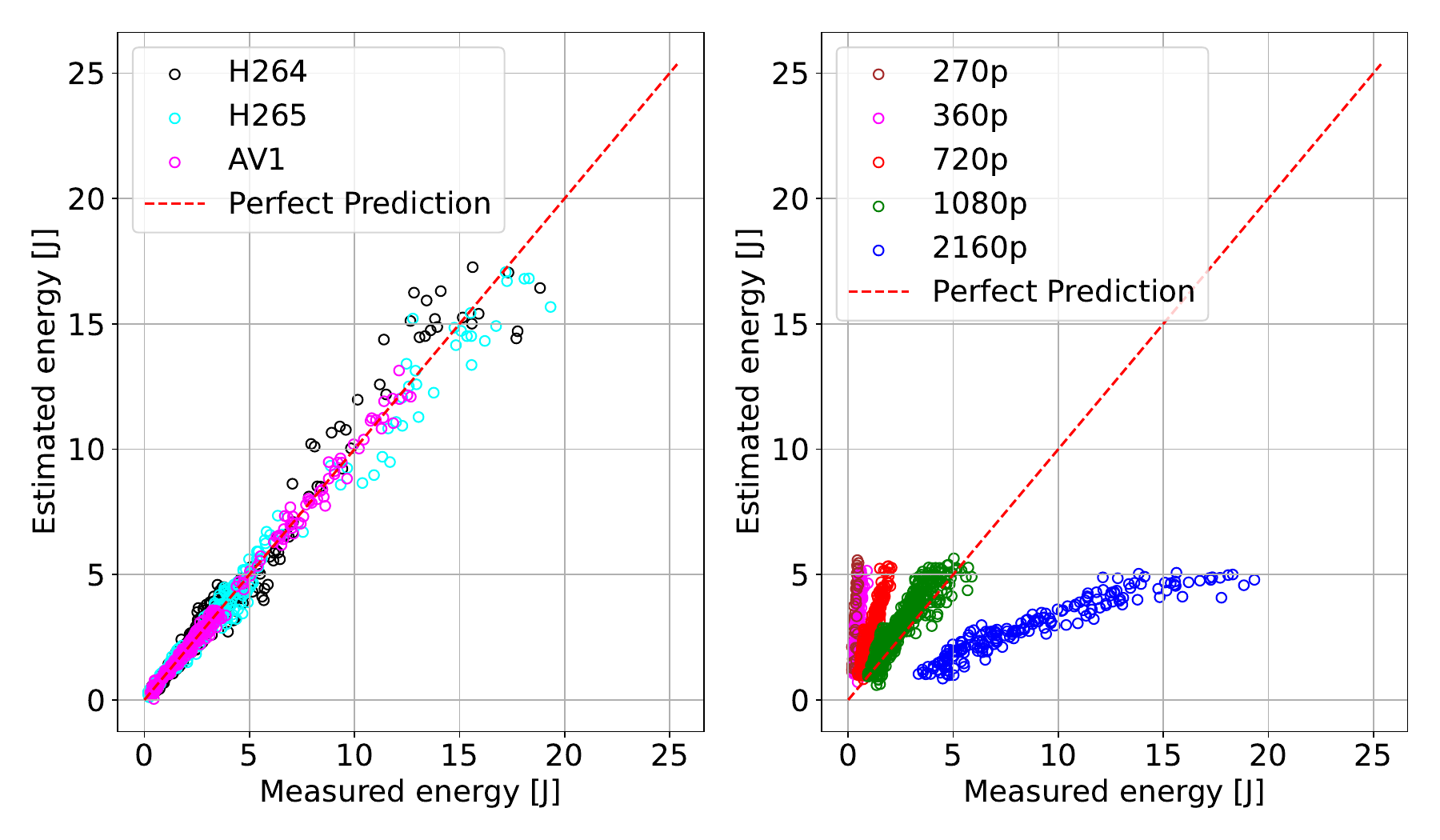}
    \caption{ Visualization of ablation study modelling results. (left) Scenario d in Table \ref{tab:AblationStudyFeatureImpact} grouped by standard. (right) Scenario a in Table \ref{tab:AblationStudyFeatureImpact} grouped by vertical spatial resolution.}
    \label{fig:AblationStudyFeatureImpactModellingResults}
\end{figure}

\begin{table}
\centering
\caption{MAPE value when a feature is set to a constant value}
\begin{tabular}{C{2cm}|C{4cm}|C{2cm}} % Defines column widths
\hline
\textbf{Scenario} & \textbf{Removed Feature} & \textbf{MAPE (in \%)} \\
\hline
a & number of pixels (width $\times$ height) & 164.70 \\
b & preset info & 37.38 \\
c & number of encoded frames & 17.43 \\
d & standard info & 10.25 \\
e & QP & 8.74 \\ 
\end{tabular}
\label{tab:AblationStudyFeatureImpact}
\end{table}

\subsection{Ablation Study}
To test the impact of each feature on the accuracy, we performed energy estimation by removing a feature. In practice, this is achieved by setting the particular feature to a constant value. Table \ref{tab:AblationStudyFeatureImpact} shows the results of this experiment. Offset energy feature in Table \ref{tab:HLFeatureList} is always a constant value, hence it is not considered in this study. Standard info in Table \ref{tab:AblationStudyFeatureImpact} corresponds to a scenario when features $\mathrm{x}_3$, $\mathrm{x}_4$, and $\mathrm{x}_5$ in Table \ref{tab:HLFeatureList} are all set to one. Similarly, the preset info refers to the case when $\mathrm{x}_6$ and $\mathrm{x}_7$ are set to one. Table \ref{tab:AblationStudyFeatureImpact} indicates that the number of pixels (or resolution) feature has the highest impact on prediction accuracy, followed by the preset information and the number of frames feature. The table also demonstrates that coding standard information has a limited impact and furthermore, it also illustrates that deletion of QP information improves the accuracy marginally. A potential explanation is the inconsistent relationship between QP value and energy observed in Fig. \ref{fig:EnergyvsQP}. 

In Fig. \ref{fig:AblationStudyFeatureImpactModellingResults}, we examine the estimation results for two scenarios, namely a and d, one with the highest impact on the accuracy and the other with a limited effect. Comparison of the results in the left plot of Fig. \ref{fig:ModellingResults} and the left plot of Fig. \ref{fig:AblationStudyFeatureImpactModellingResults} exhibit a negligible difference which explains the reason for a minor increase in MAPE for scenario d, however comparison of the right plots in the same two figures show a major difference in the prediction results. It demonstrates that spatial resolution is a key feature for modelling the energy of a HW encoder. 
\begin{table}
\centering
\caption{Accuracy with different modelling types}
\begin{tabular}{C{4cm}|C{2cm}} % Defines column widths
\hline
\textbf{Model} & \textbf{MAPE (in \%)} \\
\hline
GPR & 9.08 \\
LR & 72.98 \\
\end{tabular}
\label{tab:AblationStudyGPRvsLR}
\end{table}

\begin{figure}
    \includegraphics[width=0.5\textwidth]{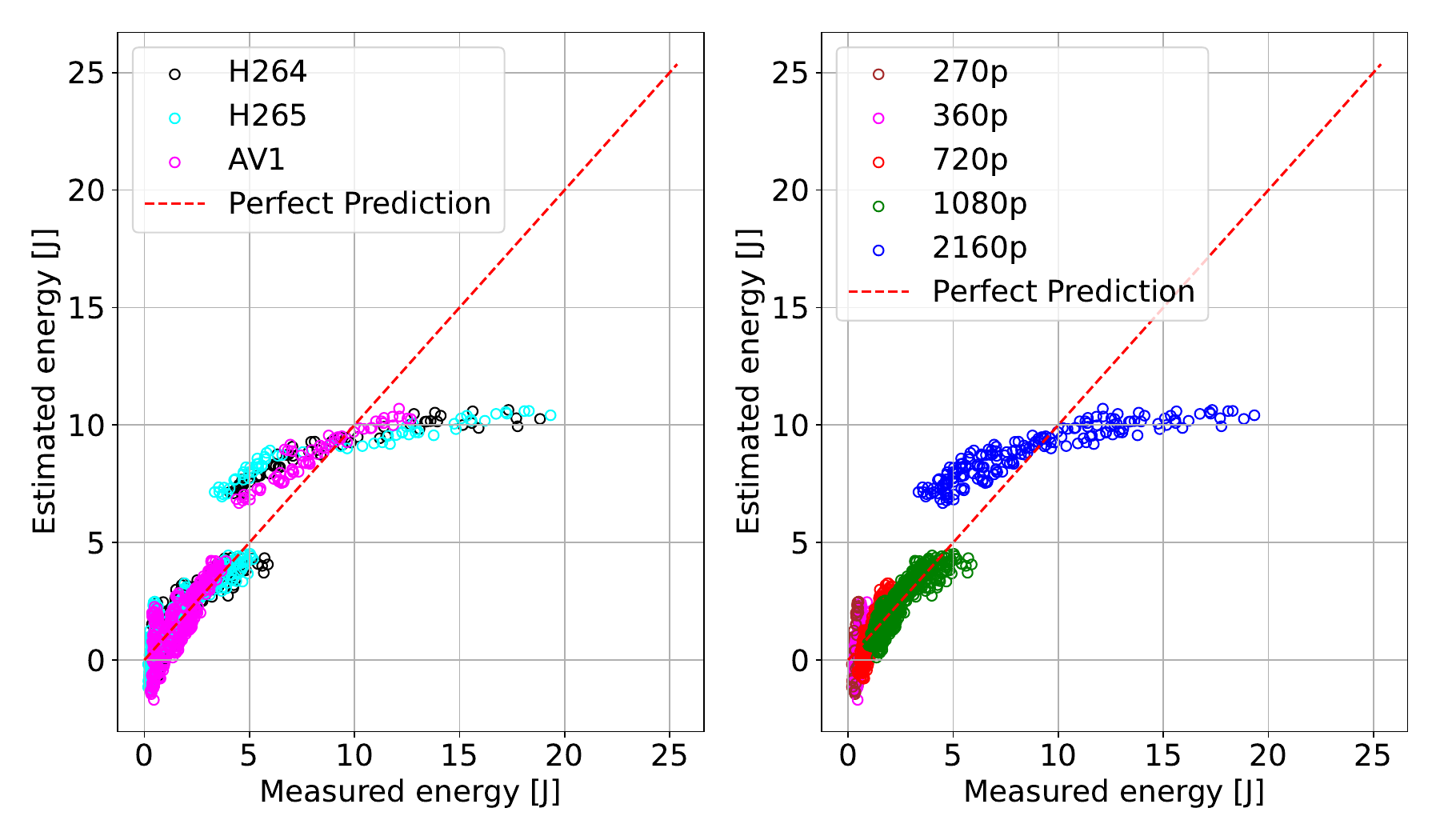}
    \caption{Visualization of prediction results with LR modelling. (left) Grouped by coding standard. (right) Grouped by vertical spatial resolution.}
    \label{fig:ModellingResultsLR}
\end{figure}

We also tested the encoder energy estimation accuracy with a LR model with features listed in Table \ref{tab:HLFeatureList}. LR is a linear model and more intuitive compared to our GPR model. As shown in Table \ref{tab:AblationStudyGPRvsLR}, we observed a MAPE of 72.98\% with the LR model, which is considerably higher than our GPR model. Results in Fig. \ref{fig:ModellingResultsLR} indicate that a LR model is not sufficient to capture the characteristics of encoding energy demand in our case, and hence, inadequate for HW encoding energy prediction.

\section{Conclusion}
\label{sec:Conclusion}
This paper introduces a HL feature model built on GPR that predicts the HW video encoding energy with a MAPE of $\sim$9\%. Furthermore, it examines the impact of each HL feature on the estimation accuracy. Finally, it presents evidence corroborating previous findings that a GPR model outperforms a LR model at energy demand prediction. The HL model in this paper does not consider video content-related features, which could further improve prediction accuracy. Elaborating on the findings in Section \ref{sec:Results}, a comprehensive analysis of the encoding energy consumption of HW and SW video encoders spanning multiple standards, while accounting for rate and distortion is an interesting topic for future work.

\bibliographystyle{IEEEtran}
\bibliography{BibFiles/IEEEReferences}
\vspace{12pt}

\end{document}